\documentclass[11pt]{article}
\usepackage{amssymb,amsmath}
\newtheorem{thm}{Theorem}
\newtheorem{cor}[thm]{Corollary}
\newtheorem{prop}[thm]{Proposition}
\newtheorem{lemma}[thm]{Lemma}

\newcommand {\bC} {\mathbb {C}}
\newcommand {\bR} {\mathbb {R}}
\newcommand {\bN} {\mathbb {N}}
\newcommand {\bT} {\mathbb {T}}
\newcommand {\bZ} {\mathbb {Z}}

\newcommand {\cA} {\mathcal {A}}
\newcommand {\cH} {\mathcal {H}}
\newcommand {\cK} {\mathcal {K}}

\newcommand {\pa} {\partial}

\def\mapright#1{\smash{\mathop{\longrightarrow}\limits^{#1}}}
\newcommand {\proof} {\noindent{\it Proof. }}

\begin{document}

\title{Ergodic properties of the quantum geodesic flow on tori}

\author{\\ S\l awomir Klimek \\ Department of Mathematics \\
Indiana University Purdue University Indianapolis\\ 402 N. Blackford St.
\\ Indianapolis, IN 46202 USA \\ \\ \\
Witold Kondracki \\ Institute of Mathematics \\ Polish Academy of Sciences\\ 
Ul. Sniadeckich 8 \\ Warsaw, POLAND}
\date{December 20, 2003}

\maketitle

\begin{abstract}
\vskip 0.3cm
\noindent We study ergodic averages for a class of pseudodifferential operators on the flat N-dimensional torus with respect to the Schr\"odinger evolution. The later can be consider a quantization of the geodesic flow on $\bT^N$. We prove that, up to semi-classically negligible corrections, such ergodic averages are translationally invariant operators.
\end{abstract}

\section{Introduction}\label{intesec}

In this paper we study time averages 
for a class of pseudodifferential operators on 
the $N$- dimensional torus $\bT^N=\bR^N/(2\pi\bZ)^N$ equipped with the 
natural flat Riemannian metric. More generally, instead of the torus one can
consider a compact manifold $X$ and a Riemannian metric $g$ on $X$.
Denote by $\Delta$ the corresponding Laplace operator (on functions).

Let $Q$ be a quantization  of the cotangent space $T^*X$ of $X$, i.e.
a linear correspondence $F\to Q(F)$, that  to a class of 
functions $F:T^*X\to \bC$ associates operators $Q(F)$ in $L^2(X,d\mu)$. Let $\cA$ be an algebra generated by all the operators $Q(F)$.
The most popular choices for $Q(F)$ are pseudo-differential operators on $X$ with symbols $F$. 

The averages are taken with respect to the time evolution given by a unitary
group $\left\{e^{itH}\right\}$ of operators in $L^2(X,d\mu)$.
There are two common choices for the generator H of the evolution:
$H=\sqrt{-\Delta}$ or $H=-\frac{1}{2}\Delta$. The later is the choice in this paper.
This operator has discrete spectrum,
$H\phi_n=\mu_n\phi_n$.
We can arrange $\mu_n$ so that $\mu_1\leq\mu_2\ldots\to\infty$.

The ergodic averages are, by definition, the following expressions:
\begin{eqnarray}
<Q(F)>:=\lim\limits_{T\to\infty}{1\over T}
\int_0^Te^{-itH}Q(F)e^{itH}\,dt.\label{averref}
\end{eqnarray}
The problem considered in this paper is to compute such averages and in particular to relate them to the geodesic flow on $T^*X$.

This problem belongs to the area of quantum chaos, which
studies semiclassical asymptotics of eigenfunctions and eigenvalues of quantum systems.
Here $H$ is the quantum hamiltonian of the free particle on $X$, and can be considered
a quantization of the geodesic flow on $T^*X$. Detailed survey of this area is contained 
in \cite{Z3} covering both the motivation and the overview of the results.

One of the main problems in that theory is to study possible limit points of 
the sequence of functionals
\begin{eqnarray}
\{F\to(\phi_n,Q(F)\phi_n)\}.\label{seqref}
\end{eqnarray}
The point is that the ergodic averages (\ref{averref}) may
be more tractable than general operators while not changing the limit structure
of (\ref{seqref}) since:
\begin{eqnarray*}
(\phi_n,Q(F)\phi_n)=(\phi_n,<Q(F)>\phi_n),
\end{eqnarray*}
which gives a motivation for our study. 

There are numerous results and conjectures concerning those 
limit points which depend on whether the corresponding geodesic flow is ergodic or not.
The case of ergodic geodesic flow is more interesting and will be briefly reviewed in
Section \ref{semisec}.
If the geodesic flow is not ergodic but completely integrable,
as is the case for $\bT^N$ considered in this paper, the limits of (\ref{seqref})
are complicated, possibly reflecting 
different ergodic components (invariant tori) of the geodesic flow. Some results in that direction are described in \cite{Z2} and \cite{J}.
The main outcome of our analysis is that the ergodic averages are, up to semiclassically
negligible correction, equal to their classical averages:
\begin{eqnarray*}
<Q(F)>=Q(\bar F)+a_F,
\end{eqnarray*}
where $\bar F(p)={1\over (2\pi)^N}\int_{\bT^N}F(x,p)\,d^Nx$,
and $a_F$ is a semi-classically negligible operator - see below. Moreover, if N=1, then
$a_F$ is a compact operator.

The paper is organized as follows. In section \ref{semisec} we shortly discuss ergodic averages in different situations. Section \ref{opersec} describes the class of 
pseudodifferential operators used for our analysis. In section \ref{ergodsec} we study 
the ergodic averages.
That section contains the main result.

\section{Semi-classical Ergodicity}\label{semisec}

In what follows we give a short overview of quantum ergodic theorems.
Let $\cA$ be an algebra of operators acting on a Hilbert space $\cH$, and let
$\rho_t$ be a one-parameter group of $*$-automorphisms of $\cA$ of the
form
\begin{eqnarray*}
\rho_t(a)=e^{itH}ae^{-itH},\ \ \ a\in\cA ,
\end{eqnarray*}
where $H$ is a self-adjoint, usually unbounded operator on $\cH$.
The main object of ergodic theory are time averages
\begin{eqnarray*}
<a>:=\lim_{T\to\infty}{1\over T}\int_0^T\rho_t(a)\,dt,
\end{eqnarray*}
where the limit above is taken with respect to a suitable
topology. Under convenient assumptions, this
limit exists and is $\rho_t$-invariant. The dynamical system
$(\cA,\rho_t)$ is then called {\it ergodic} if the following statement
holds (ergodic theorem):
\begin{eqnarray*}
<a>=\tau(a)I\,
\end{eqnarray*}
where $\tau$ is a $\rho_t$- invariant state on $\cA$. 
For examples of this situation see \cite{KL2},
\cite{KL3}, \cite{KLMR}.

In many important quantum mechanical problems this scenario does not apply
\cite{Z3}, \cite{KL1}. Let
us assume that the spectrum of $H$ is purely discrete:
\begin{eqnarray*}
H\phi_n=\mu_n\phi_n,
\end{eqnarray*}
and $\mu_n\to\infty$.
One considers the following special, invariant
state on $\cA$:
\begin{eqnarray}
\tau(a)=\lim_{E\to\infty}\tau_E(a):=\lim_{E\to\infty}{1\over\#\{\mu_n\leq
E\}}\sum_{\mu_n\leq E} (\phi_n,a\phi_n).\label{stateref}
\end{eqnarray}
Intuitively this state captures the information from very large eigenvalues
of $H$ (=high energies)
and is usually taken as the starting point of the semiclassical ergodic theory.
An operator $a$ is called semi-classically negligible if
\begin{eqnarray*}
\tau(a^*a):=\lim_{E\to\infty}\tau_E(a^*a)=0.
\end{eqnarray*}
The dynamical system $(\cA,\rho_t)$ is then called {\it semiclassically ergodic} 
if the following statement holds (semiclassical ergodic theorem):
\begin{eqnarray}
<a>=\tau(a)I+C_a,
\label{semiergref}
\end{eqnarray}
where $C_a$ is semi-classically negligible.
A theorem of Zelditch \cite{Z1} gives
sufficient conditions for (\ref{semiergref}) to hold. Let $\pi_\tau$ 
be the GNS representation of $\cA$ associated with $\tau$,
where $\tau$ is given by (\ref{stateref}). The theorem says that if we
assume that the algebra $\pi_\tau(\cA)$ is commutative, and 
that $\pi_\tau(\rho_t)$ is ergodic (in the classical sense), then the system
$(\cA,\rho_t,\tau)$ is semiclassically ergodic i.e. 
(\ref{semiergref}) holds.

We now describe the main example of a semiclassically ergodic system.
Let $Q(f)$ be a $\psi$DO of order zero on $X$ with symbol
$f:T^*X\mapsto \bC$. Let $f_P$ be the principal symbol of $Q(f)$.
It is a homogeneous function on the cotangent bundle so, effectively,
it is a function on the unit sphere bundle of the cotangent bundle 
$f_P:ST^*X\mapsto\bC$.
Let $\gamma_t$ be the geodesic flow on $ST^*X$ and denote by $dL$ the
Liouville measure on $ST^*X$. This measure is invariant with respect to
the geodesic flow $\gamma_t$. Moreover, if $(X,g)$ has constant negative 
curvature then $\gamma_t$ is ergodic.

Take $H:=\sqrt{-\Delta}$.
Using $\psi$DO theory one shows that
\begin{eqnarray*}
\tau(Q(f)):=\lim_{E\to\infty}{1\over\#\{\mu_n\leq
E\}}\sum_{\mu_n\leq E} (\phi_n,Q(f)\phi_n)=\int_{ST^*X}f_P\,dL.
\end{eqnarray*}
Let $\cA$ be the $\bC^*$-algebra obtained as the norm closure of
the algebra of order zero $\psi$DO on $X$. It is well known, and not
difficult to prove, that $\cA$ has the following structure:
\begin{eqnarray*}
0\quad\mapright{}\quad{\cK}\quad\mapright{}\quad
{\cA}\quad\mapright{\sigma}\quad C(ST^*X)\quad\mapright{}\quad0.
\end{eqnarray*}
Here $\cK$ is the ideal of compact operators in $L^2(X,d\mu)$ and
$\sigma$ is called the symbol map. It follows that if $\pi_\tau$ the GNS
representation of $\cA$ associated with $\tau$, we have $\pi_\tau(\cA)=
C(ST^*X)$, which is abelian. The Egorov's theorem implies now that
$\pi_\tau(\rho_t)=\gamma_t$, the geodesic flow on $ST^*X$. So, if $X$
has constant negative curvature, the ergodic theorem (\ref{semiergref}) follows.
From here it is not difficult to see that the following theorem, due 
to Colin de Verdiere \cite{CdV} and Schnirelman \cite{S}, is true:
if $X$ has constant negative curvature, there is a subset
$S\subset \bN$ of density $1$ such that
\begin{eqnarray}
\lim_{n\in S,n\to\infty}(\phi_n,Q(f)\phi_n)=\int_{ST^*X}f_P\,dL.
\label{densityref}
\end{eqnarray}
It was conjectured in \cite{RS} that the remainder term
$C_a$ in (\ref{semiergref})  is compact so that the limit in (\ref{densityref}) exists
for every subsequence. If true, this would imply that no
``scars" exist for such systems.

Finally, when $X=\bT^N$, the geodesic flow is not ergodic but is completely integrable and the corresponding quantum system is not semiclassically ergodic.
The main result of this paper establishes an analog of (\ref{semiergref})
in this case.

\section{Operators}\label{opersec}

First we introduce a class of pseudodifferential operators suitable for our analysis.
In what follows $\bT^N$ is the $N$- dimensional torus considered as the quotient
$\bT^N=\bR^N/(2\pi\bZ)^N$, and equipped with the induced flat Riemannian metric.
The cotangent space $T^*(\bT^N)$ of $\bT^N$ is naturally identified with
$\bT^N\times\bR^N$. The operators we are interested in are constructed using
symbols, which are functions on $\bT^N\times\bR^N$. To specify the class
of functions we need the following notation: for a  bounded function $F$ on $\bT^N\times\bR^N$,
continuous in the $\bT^N$ direction,
let $\widehat F$ be its partial Fourier transform along $\bT^N$ i.e.
\begin{eqnarray*}
\widehat F(k,p)={1\over (2\pi)^N}\int_{\bT^N}F(x,p)e^{-ikx}\,d^Nx.
\end{eqnarray*}
Define $C_r(\bT^N\times\bR^N)$, $r>N$, to be the space of functions $F$ on $\bT^N\times\bR^N$
such that
\begin{eqnarray}
||F||_r:={\mathop{\sup}_{(k,p)\in\bZ^N\times\bZ^N}}
\left|(1+|k|^2)^{r/2}\,\widehat F(k,p)\right|<\infty.
\label{normdefref}
\end{eqnarray}
It should be noted that in the above definition only the values of $F(x,p)$ on $\bT^N\times\bZ^N$
matter as justified below by the formula (\ref{qfdefref}) for the operator $Q(F)$ defined by $F$.
Also $C_r(\bT^N\times\bR^N)$ is a Banach space and 
linear combinations of functions of the form $f(x)g(p)$, where $f$ is a trigonometric polynomial,
form a dense subspace in $C_r(\bT^N\times\bR^N)$. Moreover we have:

\begin{prop}\label{normprop}
\begin{eqnarray*}
{1\over C_r}\ {\mathop{\sup}_{(x,p)\in\bT^N\times\bZ^N}}|F(x,p)|\leq||F||_r\leq
{\mathop{\sup}_{(x,p)\in\bT^N\times\bZ^N}}|(1-\Delta)^{r/2}F(x,p)|
\end{eqnarray*}
where $\Delta$ is the laplacian on $\bT^N$ and
\begin{eqnarray}
C_r=\sum_{k\in\bZ^N}\left(1+|k|^2\right)^{-r/2}.\label{crdefref}
\end{eqnarray}
(As stated above we assume $r>N$ throughout the paper).
\end{prop}
\proof  We have:
\begin{eqnarray*}
&&{\mathop{\sup}_{(x,p)\in\bT^N\times\bZ^N}}|F(x,p)|={\mathop{\sup}_{(x,p)\in\bZ^N\times\bZ^N}}
\left|\sum_{k\in\bZ^N}\widehat F(k,p)e^{ikx}\right|\leq\\ &&\leq{\mathop{\sup}_{p\in\bZ^N}}
\left|\sum_{k\in\bZ^N}\widehat F(k,p)\right|\leq\\
&&\leq{\mathop{\sup}_{(k,p)\in\bZ^N\times\bZ^N}}\left|(1+|k|^2)^{r/2}\,\widehat F(k,p)\right|\ 
\sum_{k\in\bZ^N}\left(1+|k|^2\right)^{-r/2}.\\
\end{eqnarray*}
which proves the first inequality. On the other hand we have:
\begin{eqnarray*}
&&{\mathop{\sup}_{(k,p)\in\bZ^N\times\bZ^N}}\left|(1+|k|^2)^{r/2}\,\widehat F(k,p)\right|
={\mathop{\sup}_{(k,p)\in\bZ^N\times\bZ^N}}|\widehat{(1-\Delta)^{r/2}F(k,p)}|\leq\\
&&\leq{\mathop{\sup}_{(x,p)\in\bT^N\times\bZ^N}}|(1-\Delta)^{r/2}F(x,p)|,
\end{eqnarray*}
because $|\widehat f(k)|\leq{\mathop{\sup}_{x\in\bT^N}}|f(x)|$.
$\square$

\medskip
To each symbol $F\in C_r(\bT^N\times\bR^N)$ we associate a bounded operator $Q(F)$
in $L^2(\bT^N,d^Nx)$ by the following, global version of the usual formula:
\begin{eqnarray*}
Q(F)\psi(x)={1\over (2\pi)^N}\sum_{k\in\bZ^N}\int_{\bT^N}F(x,k)\,e^{ik(x-y)}\,\psi(y)\,d^Ny.
\end{eqnarray*}
Notice also that writing $\psi\in L^2(\bT^N,d^Nx)$ in the Fourier series:
$$\psi(x)=
\sum_{k\in\bZ^N}\psi_ke^{ikx},
$$
we have
\begin{eqnarray}
Q(F)\psi(x)=\sum_{k\in\bZ^N}\psi_k\,F(x,k)\,e^{ikx}.\label{qfdefref}
\end{eqnarray}

\begin{lemma}\label{normlemma}
We have
\begin{eqnarray*}
||Q(F)||\leq C_r||F||_r,
\end{eqnarray*}
where $C_r$ is given by (\ref{crdefref}).
\end{lemma}
\proof Note that in the formulas below the summation indexes run over $\bZ^N$.
We study $(\psi,Q(F)\psi)$ with $\psi(x)=\sum\psi_ke^{ikx}$. Using (\ref{qfdefref})
we obtain:
\begin{eqnarray*}
&&(\psi,Q(F)\psi)=\sum_{k,l,m}\bar\psi_k\psi_m\widehat F(l,m)\,\int_{\bT^N}
e^{i(l+m-k)x}\,
{d^Nx\over (2\pi)^N}=\\
&&=\sum_{k,m}\bar\psi_k\psi_m\widehat F(k-m,m)=
\sum_{k,m}\bar\psi_k\psi_m\widehat F(k-m,m){(1+(k-m)^2)^{r/2}\over (1+(k-m)^2)^{r/2}}.
\end{eqnarray*}
It follows that:
\begin{eqnarray*}
&&|(\psi,Q(F)\psi)|\leq ||F||_r\,\sum_{k,m}|\psi_k|\,|\psi_m|\,
(1+(k-m)^2)^{-r/2}=\\
&&= ||F||_r\,\sum_{k,l,m}|\psi_k|\,|\psi_m|\,(1+l^2)^{-r/2}\,
\int_{\bT^N}e^{i(l+m-k)x}\,
{d^Nx\over (2\pi)^N}
\end{eqnarray*}
Let $\tilde\psi(x)=\sum_k|\psi_k|e^{ikx}$. Notice that $||\tilde\psi||=||\psi||$.
We can use $\tilde\psi$ to write the above estimate as
\begin{eqnarray*}
&&|(\psi,Q(F)\psi)|\leq ||F||_r\,(\tilde\psi,\sum_l(1+l^2)^{-r/2}\,e^{ilx}\ \tilde\psi)\leq\\
&&\leq ||F||_r\,{\mathop{\sup}_{x\in\bT^N}}\left|\sum_l(1+l^2)^{-r/2}\,e^{ilx}\right|
\,||\tilde\psi||^2\leq C_r\,||F||_r\,||\psi||^2,
\end{eqnarray*}
and the claim follows.
$\square$

\begin{prop}\label{qcompprop}
If $F\in C_r(\bT^N\times\bR^N)$ and $|F(x,p)|\to 0$ as $p\to \infty$,
then $Q(F)$ is a compact operator.
\end{prop}
\proof  First notice that if $|F(x,p)|\to 0$ as $p\to \infty$ then also
$|\widehat F(k,p)|\to 0$ as $p\to \infty$. Indeed:
\begin{eqnarray*}
|\widehat F(k,p)|=\left|{1\over (2\pi)^N}\int_{\bT^N}F(x,p)e^{-ikx}\,d^Nx\right|
\leq {\mathop{\sup}_{x\in\bT^N}}|F(x,p)|\to 0.
\end{eqnarray*}
Consider now $F_L(x,p):=\sum_{|k|\leq L}\widehat F(k,p)e^{ikx}$. It follows that
$|F_L(x,p)|\to 0$ as $p\to \infty$. Additionally $F_L\to F$ in $C_r(\bT^N\times\bR^N)$, which
by Lemma \ref{normlemma} implies $Q(F_L)\to Q(F)$ in norm.
Finally $Q(F_L)$ is a compact operator. Indeed, for each Fourier coefficient $\widehat F(k,p)$
we have
\begin{eqnarray*}
Q(\widehat F(k,\cdot))e^{inx}=F(k,n)e^{inx}.
\end{eqnarray*}
This means that $Q(\widehat F(k,\cdot)$ has discrete spectrum equal to $F(k,n)$,
$n\in\bZ^N$ which goes to 0 as $n\to\infty$, implying that $Q(\widehat F(k,\cdot)$
and also $Q(F_L)$ are compact operators. Finally, $Q(F)$ is compact as a norm
limit of compact operators.
$\square$
\medskip

The following observation is worth a notice even though it is not used in the remainder of the paper.
It gives more information on the topological properties of the quantization map
$F\to Q(F)$.

\begin{prop}
If $F_n\to F$ pointwise (a.e.) and if $\{F_n\}$ have uniformly bounded  norms (in 
$C_r(\bT^N\times\bR^N)$) then $Q(T_n)\to Q(T)$ weakly.
\end{prop}
\proof First notice that Proposition \ref{normprop} and uniform boundedness of $C_r(\bT^N\times\bR^N)$
norms implies that $\{F_n\}$ are uniformly bounded, that is, their sup norms are uniformly
bounded.
It follows using Lebesque dominated convergence theorem that 
$\widehat F_n\to\widehat F$ pointwise. To show weak convergence of  $Q(F_n)$
we study $(\psi,Q(F_n)\psi)$ just as in Lemma \ref{normlemma}. We have
\begin{eqnarray*}
(\psi,Q(F_n)\psi)=\sum_{k,m}\bar\psi_k\psi_m\widehat F_n(k-m,m),
\end{eqnarray*}
and by the previous remark the inside of the sum converges pointwise to
$\bar\psi_k\psi_m\widehat F(k-m,m)$ as $n\to \infty$. Also
\begin{eqnarray*}
&&|\bar\psi_k\psi_m\widehat F_n(k-m,m)|=\left|\bar\psi_k\psi_m\widehat 
F_n(k-m,m){(1+(k-m)^2)^
{r/2}\over (1+(k-m)^2)^{r/2}}\right|\leq\\
&&\leq (\mathop{\sup_n} ||F_n||_r)\,|\psi_k|\,|\psi_m|\,(1+(k-m)^2)^{-r/2}.
\end{eqnarray*}
The last expression is summable just as in the proof of Lemma \ref{normlemma}, so
using Lebesque dominated convergence theorem again concludes the proof.
$\square$

\section{Ergodic averages}\label{ergodsec}

Let $H$ be the following operator in $L^2(\bT^N,d^Nx)$: 
$$
H:=-{1\over 2}\Delta=-{1\over 2}\left({\pa^2\over\pa x_1}+\ldots+{\pa^2\over\pa x_N}
\right),
$$
and let $\phi_n=e^{inx}\in L^2(\bT^N,d^Nx)$. Then $H\phi_n=\mu_n\phi_n$ with $\mu_n=n^2/2$.
We are interested in studying the norm limits:
$$<Q(F)>:=\lim_{T\to\infty}<Q(F)>_T:=\lim_{T\to\infty}{1\over
T}\int_0^Te^{-itH}Q(F)e^{itH}\,dt,
$$
called ergodic averages.

The following general lemma deals with ergodic averages of compact
operators.
\begin{lemma}\label{compactlemma}
If $C$ is a compact operator and $H$ has discrete spectrum
with finite multiplicities, then $\lim\limits_{T\to\infty}{1\over
T}\int_0^Te^{-itH}Ce^{itH}\,dt$ exists in norm and is a compact operator.
\end{lemma}
\proof We study the weak limit first:
$$
<C>={\rm w}-\lim_{T\to\infty}<C>_T={\rm w-}\lim_{T\to\infty}{1\over
T}\int_0^Te^{-itH}Ce^{itH}\,dt.
$$
The matric elements with respect to the basis of eigenvectors of $H$ are
$$
(\phi_i,<C>_T\phi_j)=(\phi_i,C\phi_j){1\over T}\int_0^T
e^{it(\lambda_i-\lambda_j)}\,dt.
$$
Consequently,
\begin{eqnarray}
(\phi_i,<C>\phi_j)=\begin{cases}0& \text{if $\lambda_i=\lambda_j$}\\
(\phi_i,C\phi_j)& \text{otherwise.}\\
\end{cases} \label{mateleref}
\end{eqnarray}
We need the following notation. Writing the spectral decomposition of
the Hilbert space $\cH=\oplus_i\cH_i$ with respect to operator $H$,
denote by $P_i$ the orthogonal projection onto $\cH_i$, the eigenspace
corresponding to eigenvelue $\lambda_i$. Projections $P_i$ are mutually
orthogonal $P_iP_j=\delta_{ij}P_i$. We will also need $P_{\leq n}:=
\sum_{i\leq n}P_i$, and $P_{\geq n}:=\sum_{i\geq n}P_i$. We have
$$
P_iP_{\geq n}=P_i,
$$
if $i\geq n$. In this notation we have
$$
<C>=\sum_iP_iCP_i,
$$
where the limit defining the series is in the weak topology. We claim
that $<C>$ is compact. To prove it we show that
$$
||\sum_{i\geq n}P_iCP_i||\to 0\ \ {\rm as}\ \ n\to\infty.
$$
This implies that $<C>$ is compact as a norm limit of compact operators.
We compute:
\begin{eqnarray*}
&&||\sum_{i\geq n}P_iCP_ix||^2=\sum_{i\geq n}||P_iCP_ix||^2=
\sum_{i\geq n}||P_iP_{\geq n}CP_ix||^2\leq\\
&&\leq||P_{\geq n}C||^2\,\sum_{i\geq n}||P_ix||^2
\leq||P_{\geq n}C||^2\,||x||^2.
\\
\end{eqnarray*}
Consequently,
$$
||\sum_{i\geq n}P_iCP_i||\leq||P_{\geq n}C||\to 0\ \ {\rm as}\ \ n\to\infty,
$$
since $C$ is compact.

So far we have
$$
{\rm w}-\lim_{T\to\infty}<C>_T=<C>,
$$
with compact $<C>$. Observe also that the averaging is a contraction:
$$
||<C>_T||\leq||C||.
$$
To prove that $<C>_T$ converges to $<C>$ in norm we set up an approximation 
argument. To this end we introduce $C_n:=P_{\leq n}CP_{\leq n}$. We have:
\begin{itemize}
\item $C_n\to C$ in norm (by compactness of $C$).
\item $<C_n>_T\to <C>_n:=P_{\leq n}<C>P_{\leq n}$ weakly,
but since $C_n$ are finite dimensional, it means in norm.
\item $<C>_n\to <C>$ in norm (by compactness of $<C>$).
\end{itemize}

It follows that in the estimate
\begin{eqnarray*}
&&||<C>_T-<C>||\leq||<C-C_n>_T||+\\
&&+||<C_n>_T-<C>_n||+||<C>_n-<C>||,\\
\end{eqnarray*}
all three terms are small.
$\square$

\begin{cor}
If $F\in C_r(\bT^N\times\bR^N)$ and $|F(x,p)|\to 0$ as $p\to \infty$,
then $<Q(F)>$ is a compact operator.
\end{cor}
\proof This immediately follows from Proposition \ref{qcompprop}  and 
Lemma \ref{compactlemma}.
$\square$
\medskip

Next we study ergodic averages of arbitrary operators $Q(F)$.
To this end we obtain an explicit formula for $e^{itH}Q(F)e^{-itH}$.
\begin{lemma}\label{geoflowlemma}
If $F\in C_r(\bT^N\times\bR^N)$ then
\begin{eqnarray}
e^{itH}Q(F)e^{-itH}=Q(F_t),\label{conjref}
\end{eqnarray}
where $F_t\in C_r(\bT^N\times\bR^N)$ and
\begin{eqnarray}
F_t(x,p)=e^{-it\Delta_x/2}\left(F(x+tp,p)\right),\label{ftdefref}
\end{eqnarray}
where $\Delta_x={\pa^2\over\pa x_1}+\ldots+{\pa^2\over\pa x_N}
$ is the Laplace operator in the $x$-coordinate.
In Fourier transform:
\begin{eqnarray}
\widehat F_t(k,p)=e^{it(k^2/2+kp)}\,\widehat F(k,p).\label{ftfuref}
\end{eqnarray}
\end{lemma}

Notice that in (\ref{ftdefref}) the argument shift $x\to x+tp$ is precisely the
geodesic flow on the cotangent space of $\bT^N$ with respect to the flat metric. 
The action of $e^{-it\Delta_x/2}$ on $F$ in (\ref{ftdefref}) is a quantum effect,
and, as we will see, it does influence the ergodic properties of the operators.

\proof Since $\widehat F$ and $\widehat F_t$ differ by a phase 
it follows that $F$ and $F_t$ have the same norm:
$$||F_t||_r=||F||_r.
$$
Because of this, the usual $2\epsilon$ argument shows that it is enough to verify
(\ref{conjref}) for a dense set of $F$'s. Indeed, if $F_n\to F$ and (\ref{conjref}) holds for $F_n$
then in the following estimate all terms are small:
\begin{eqnarray*}
&&||e^{itH}Q(F)e^{-itH}-Q(F_t)||\leq ||e^{itH}(Q(F)-Q(F_n))e^{-itH}||+\\
&&+||e^{itH}Q(F_n)e^{-itH}-Q((F_n)_t)||+||Q((F_n)_t)-Q(F_t)||.
\end{eqnarray*}
Also, both sides of (\ref{conjref}) are linear in $F$ so it is enough to verify the hypothesis for functions in the following form:
$F(x,p)=e^{ikx}g(p)$, since linear combinations of them form a dense set in 
$C_r(\bT^N\times\bR^N)$. Applying $e^{itH}Q(F)e^{-itH}$ to $\psi(x)=\sum_l\psi_le^{ilx}$
we obtain
\begin{eqnarray*}
&&e^{itH}Q(F)e^{-itH}\psi(x)=\sum_le^{it(k+l)^2/2}\,e^{ikx}\,g(l)\,e^{-itl^2/2}\,\psi_l\,e^{ilx}=\\
&&=e^{itk^2/2}\sum_le^{ik(x+tl)}\,g(l)\,\psi_l\,e^{ilx}=\\
&&=Q(e^{itk^2/2}\,e^{ik(x+tp)}\,g(p))\psi(x)=Q(F_t)\psi(x),
\end{eqnarray*}
which concludes the proof.
$\square$
\medskip

Now we use Lemma \ref{geoflowlemma} to explicitly calculate ergodic averages of
$Q(F)$. It turns out
that long time averaging simply drops most frequencies out from $F$ as 
described by the following lemma.

\begin{lemma}
The norm limit $<Q(F)>:=\lim\limits_{T\to\infty}{1\over T}
\int_0^Te^{-itH}Q(F)e^{itH}\,dt$
exists and $<Q(F)>=Q(<F>)$, where 
\begin{eqnarray}
<F>=\sum_{\{k\in\bZ^N:\ kp+k^2/2=0\}}\widehat F(k,p)e^{ikx}.\label{faveref}
\end{eqnarray}
\end{lemma}
\proof
The following observations:
\begin{itemize}
\item $<\cdot>$ is a contraction: $||<Q(F)>_T||\leq||Q(F)||$,
\item norm estimate: $||Q(F)||\leq C_r||F||_r$, and
\item $<Q(F)>_T=Q\left({1\over T}\int_0^TF_t\,dt\right)$,
\end{itemize}
\noindent imply that in order to prove the lemma we need to show that the norm limit
of ${1\over T}\int_0^TF_t\,dt$ in $C_r(\bT^N\times\bR^N)$ is $<F>$. 

Let us write down the formulas 
for the Fourier components of $F_t$ and $<F>$. Formula (\ref{ftfuref}) gives:
$$
\widehat{F_t}(k,p)=e^{it(k^2/2+kp)}\,\widehat F(k,p).
$$
It follows that
$$
{1\over T}\int_0^T\widehat{F_t}(k,p)\,dt=\begin{cases}\widehat F(k,p)& \text{if $kp+k^2/2=0$}\\
{\left(e^{iT(kp+k^2/2)}-1\right)\over iT(kp+k^2/2)}\,
\widehat F(k,p)& \text{otherwise.}\\
\end{cases}
$$
Formula (\ref{faveref}) implies:
$$
\widehat{<F>}(k,p)=\begin{cases}\widehat F(k,p)& \text{if $kp+k^2/2=0$}\\
0& \text{otherwise.}\\
\end{cases}
$$
We can now estimate the norm of the difference:
\begin{eqnarray*}
&&||{1\over T}\int_0^T\widehat{F_t}\,dt-<F>||_r=\\
&&=\mathop{\sup_{\{k,p\in\bZ^N:\ kp+k^2/2=0\}}}(1+k^2)^{r/2}\,\left|
{\left(e^{iT(kp+k^2/2)}-1\right)\over iT(kp+k^2/2)}\right|\,|\widehat F(k,p)|\leq\\
&&\leq{2\over T}\,||F||_r\,\mathop{\sup_{\{k,p\in\bZ^N:\ kp+k^2/2=0\}}}
|(kp+k^2/2)^{-1}|\leq{4\over T}\,||F||_r,
\end{eqnarray*}
where in the last step we used the fact that $kp+k^2/2$ is half-integer. Clearly
this implies that ${1\over T}\int_0^TF_t\,dt$ converges in norm to $<F>$ and the thesis
follows.
$\square$
\medskip

Let, as before, $\mu_n=n^2/2$ be the eigenvalues of $H$. We use the notation 
$N(E):=\#\{\mu_n\leq E\}$
for the number of eigenvalues of $H$ which are less or equal $E$. Recall \cite{Z3} that an
operator $a$ is called semi-classically negligible (SN) with respect to H if
$$\tau(a^*a):=\lim_{E\to\infty}\tau_E(a^*a):=
\lim_{E\to\infty}{1\over N(E)}\sum_{\mu_n\leq E} (\phi_n,a^*a\phi_n)=0.
$$

\begin{prop}\label{snprop}
Semi-classically negligible operators form a closed left-sided ideal
in the algebra of all bounded operators.
\end{prop}
\proof  We need to verify the following three things:
\begin{enumerate}
\item If $a,b$ are SN so is $a+b$
\item If $a_k\to a$ and $a_k$ are SN then $a$ is also SN
\item If $a$ is SN and $b$ bounded then $ba$ is SN
\end{enumerate}

Since $(\phi_n,a^*a\phi_n)=||a\phi_n||^2$, part 1 follows from the estimate:
\begin{eqnarray}
||(a+b)\phi_n||^2\leq2(||a\phi_n||^2+||b\phi_n||^2).\label{snaddref}
\end{eqnarray}
To prove part 2 we set up a $2\epsilon$ estimate. Using (\ref{snaddref}) we get:
\begin{eqnarray*}
&&{1\over N(E)}\sum_{\mu_n\leq E}||a\phi_n||^2\leq{2\over N(E)}\sum_{\mu_n\leq E}
||(a-a_k)\phi_n||^2+\\
&&+{2\over N(E)}\sum_{\mu_n\leq E}||a_k\phi_n||^2\leq2||(a-a_k)||^2+
{2\over N(E)}\sum_{\mu_n\leq E}||a_k\phi_n||^2,
\end{eqnarray*}
and both terms are small. Finally, part 3 is obvious: $$||ba\phi_n||^2\leq||b||^2\,||a\phi_n||^2.
$$
$\square$
\medskip

The following is the main result of our paper:

\begin{thm}
With the above notation
$$<Q(F)>=Q(\bar F)+a_F,
$$
where 
$$\bar F(x,p)=\widehat F(0,p)={1\over (2\pi)^N}\int_{\bT^N}F(x,p)\,d^Nx,
$$
and $a_F$ is a semi-classically negligible operator. Moreover, if N=1, then
$a_F$ is a compact operator.
\end{thm}
\proof We can rephrase the theorem as saying that all but zero Fourier component of
$F$ give rise to operators with semi-classically negligible ergodic averages.
Also notice that typically if $\widehat F(0,p)$ does not go to 0 as $p\to\infty$ then 
$Q(\bar F)$ is not SN. 

Items 1 and 2 in Proposition \ref{snprop} let us use an approximation argument.
Indeed it is enough to verify that for
functions in the following form: $F(x,p)=e^{ikx}g(p)$, $k\neq 0$,
the corresponding ergodic averages $<Q(F)>=Q(<F>)$ are SN.

It follows from (\ref{faveref}) that for such $F$ we have
$$
<F>(x,p)=\begin{cases}e^{ikx}g(p)& \text{if $kp+k^2/2=0$}\\
0& \text{otherwise.}\\
\end{cases}
$$
Since $Q(<F>)\phi_n(x)=<F>(x,n)e^{inx}$ we have
\begin{eqnarray}
Q(<F>)\phi_n(x)=\begin{cases}e^{i(k+n)x}g(n)&\text{ if $kn+k^2/2=0$}\\
0& \text{otherwise.}\\
\end{cases}
\label{qaveref}
\end{eqnarray}
We can now estimate $\tau_E(Q(<F>)^*Q(<F>))$ as follows:
\begin{eqnarray*}
&&{1\over N(E)}\sum_{\mu_n\leq E}||Q(<F>)\phi_n||^2=\\
&&={1\over N(E)}\sum_{\{n\in\bZ^N:\ \mu_n\leq E,\ kn+k^2/2=0\}}|g(n)|^2\leq\\
&&\leq ||F||_r{\#\{n\in\bZ^N:\ {1\over2}n^2\leq E,\ kn+k^2/2=0\}
\over\#\{n\in\bZ^N:\ {1\over2}n^2\leq E\}}.
\end{eqnarray*}
In the above expression the denominator $\#\{n\in\bZ^N:\ {1\over2}n^2\leq E\}$
behaves like $E^{N/2}$ for large $E$, while the numerator, because of the constraint
$kn+k^2/2=0$ grows at most like $E^{N/2-1}$, if $N>1$. Consequently,
$$
{1\over N(E)}\sum_{\mu_n\leq E}||Q(<F>)\phi_n||^2={\rm O}(1/E),
$$
which proves that $Q(<F>)$ is SN. Consider now the case of $N=1$. We can again
use an approximation argument and, just as before, it is enough to verify that for
functions in the following form: $F(x,p)=e^{ikx}g(p)$, $k\neq 0$,
the corresponding ergodic averages $<Q(F)>=Q(<F>)$ are compact.
Examining formula (\ref{qaveref}) we see that $Q(<F>)\phi_n(x)$ is nonzero only
when $ kn+k^2/2=0$. But this equation has no solutions if $k$ is odd,
and one solution $n=-k/2$ if $k$ is even. In any case we obtain a finite
dimensional operator. Since norm limits of finite dimensional operators are
compact, the theorem is proved.
$\square$


\end{document}